\documentclass[10pt,amstex]{article}

\usepackage[usenames]{color}

\usepackage{amscd}
\usepackage{amsmath}
\usepackage{amssymb}
\usepackage{graphicx}

\newtheorem {theo} {\bf Theorem} [section]
\newtheorem {prop} [theo] {\bf Proposition}

\newtheorem {lem} [theo] {\bf Lemma}
\newtheorem {defn} [theo] {\bf Definition}

\newtheorem {rem} [theo] {\bf Remark}

\newcommand{\qed}{\nopagebreak\hfill{\vrule width6pt height6pt depth0pt}}

\newcommand{\be}{\begin{eqnarray}}
\newcommand{\ee}{\end{eqnarray}}
\newcommand{\benn}{\begin{eqnarray*}}
\newcommand{\eenn}{\end{eqnarray*}}
\newcommand{\bse}{\begin{equation}}
\newcommand{\ese}{\end{equation}}
\newcommand{\bsenn}{\begin{displaymath}}
\newcommand{\esenn}{\end{displaymath}}
\newcommand{\logand}{\;\;{\rm and }\;\;}

\newcommand{\logor}{\;\;{\rm or }\;\;}

\usepackage{graphicx}

\begin{document}

\title{Symmetry and Stability of Homogeneous Flocks\\
A Position Paper}
\author{ J. J. P. Veerman\thanks{e-mail: veerman@pdx.edu}\\
Fariborz Maseeh Dept. of Math and Stat, Portland State
Univ., Portland, OR 97201, USA.
\date{}
}\maketitle

\begin{abstract}
The study of the movement of flocks, whether biological or technological
is motivated by the desire to understand the capability of coherent motion
of a large number
of agents that only receive very limited information. In a biological flock
a large group of animals seek their course while moving in a
more or less fixed formation. It seems reasonable that the immediate
course is determined by leaders at the boundary of the flock. The others
follow: what is their algorithm? The most popular technological application
consists of cars on a one-lane road. The light turns green and the lead car
accelerates. What is the efficient algorithm for the others to closely follow
without accidents? In this position paper we present
some general questions from a more fundamental point of view.
We believe that the time is right to solve
many of these questions: they are within our reach.

\begin{center}
{\bf Keywords}
\end{center}
Coupled Oscillators, Network, Stability, Symmetry, Decentralized Control
\end{abstract}

\begin{centering}\section{Introduction} \label{chap:one}\end{centering}
\setcounter{figure}{0} \setcounter{equation}{0}

We model agents and their interactions as a \emph{finite} network of
(coupled) oscillators in $\mathbb{R}^d$. We assume that
agents observe relative positions and velocities only (no third and higher derivatives).
The best studied example of such a flock is a class of well-known models
for cars equipped with sensors and on automatic pilot that are moving in close
formation on a one-lane road. However, we are interested in a wider class
of flocks (especially those moving in $\mathbb{R}^2$ and $\mathbb{R}^3$).
Our main concerns in this position paper fall in two categories: \\
 - What are the equations we should study?\\
 - What does it mean for a (large) flock to be stable?\\
These are the questions that motivate this position paper.In our view their
will give rise to exciting new mathematics with ample implications for the sciences.

In response to the first concern we present physical considerations
that impel us to consider translational symmetry of the equations of motion
(in Section \ref{chap:two}) and rotational symmetry and its breaking
(in Section \ref{chap:four}).
These considerations are perhaps not fundamentally new:
but the literature favors detailed studies that are of
immediate technological importance. However, we argue that that
these equations can --- and should --- successfully be attacked.

The second concern is addressed in Section \ref{chap:five}.
We'll see that even in an asymptotically stable system, fluctuations may
grow very large before they attenuate and die out. In one dimension they lead to collisions.
In dimension greater than one, collisions are perhaps unlikely, but large
fluctuations could lead to loss of cohesion of the flock (because the agents can't
``see" each other anymore). Clearly there is a limit to the size
of the fluctuations a flock can sustain. We will argue that it is useful to study
this kind of stability in the time domain. We also argue that more general methods to investigate
this kind of ``transient (in)stability" should be researched.

We'll use a very simple set-up of a 1 dimensional flocks as our `standard example'.
\vskip 0.1in
\begin{defn} \emph{(Standard Example)}
Consider $N+1$ agents in $R$ labeled
from 0 to $N$, each of whose coordinates are given by $z_k(t)$ as function of time.
For $k\in \{1\cdots, N-1\}$, let $f$ and $g$ be negative and
$\rho$ and $r$ are in $(0,1)$.
\bsenn
\ddot z_k = f \left\{z_{k}-(1-\rho) z_{k-1}-\rho z_{k+1}\right\} +
   g\{\dot z_k- (1-r) \dot z_{k-1}- r \dot z_{k+1}\}
\esenn
The \emph{boundary} conditions are:
\bsenn
\ddot z_0 = 0  \quad \logand \quad
\ddot z_N = f \left\{z_{N}-z_{N-1}\right\} + g\{\dot z_N-\dot z_{N-1}\}
\esenn
The initial conditions are:
\bsenn
\forall \; k:\; z_k(0)=0\quad, \quad  \forall \; k>0: \; \dot z_k(0)=-0.1 \quad ,\quad
\dot z_0(0)=0 \quad .
\esenn
It should be understood that the original coordinates of the agents given by $x_k$
where $x_k=z_k+h_k$; those are the coordinates we display in Figure \ref{fig:TIME}.
The $\{h_k\}$ are priori chosen constants and their differences
determine the preferred \emph{relative} positions of the agents.
\label{def:standard example}
\end{defn}

\begin{centering}\section{Truly Decentralized Linear Flocks} \label{chap:two}\end{centering}
\setcounter{figure}{0} \setcounter{equation}{0}

Ever since the inception (\cite{HM1}, \cite{HM2}) of the subject it has been
a challenge to mathematically express the notion ``decentralized" (\cite{Chu}).
Here is how we interpret it: In a decentralized flock the only information an agent
receives are the position and velocity relative to it of some nearby agents.
The agents do not receive information from an outside source.
The only exceptions to this are some agents, called \emph{leaders}, who may undergo
an external forcing.

\vskip .1in
\begin{prop} \emph{(Galilean Invariance)} In the absence of external forcing, the motion of a truly decentralized
linear flock in $\mathbb{R}$ is described by ($k$ and $i$ in $\{0\cdots N\}$):
\bsenn
\ddot x_k = f \sum_{i\in{\cal N}_\rho(k)}\,L_{\rho,ki} (x_{i}-h_i) +
                    g \sum_{i\in{\cal N}_r(k)}\,L_{r,ki} \dot x_{i} \quad \logor
\quad  \ddot z_k = f \sum_{i\in{\cal N}_\rho(k)}\,L_{\rho,ki} z_{i} +
                    g \sum_{i\in{\cal N}_r(k)}\,L_{r,ki} \dot z_{i} \quad .
\esenn
where $z_k=x_k-h_k$. The row-sums of the $N+1$ by $N+1$ ``Laplacian" matrices
$L_\rho$ and
$L_r$ must be zero. The sums are over the set of agents ${\cal N}_\rho$ whose relative
velocity can be measured by $k$, resp. whose relative position can be measured
by $K$ (in the case of ${\cal N}_r$).
We furthermore have that there is a 2-parameter set of solutions
given by (for any $Z$ and $V$ in $\mathbb{R}$):
\bsenn
 \forall \; k\in \{0 \cdots, N\}\;:\; z_k(t)= Vt+Z \quad .
\esenn
\label{prop:Galilean Invariance}
\end{prop}

From now on we will always assume that $f$ and $g$ are chosen so that
the system is asymptotically stable, that is: in the absence of external forcing,
the system always converges to an orbit of the form $z_k(t)= Vt+Z$.
We call such a system \emph{stabilized}. In the standard example,
the stipulation that $f$ and $g$ be negative guarantees this. For general Laplacians,
and especially in dimension bigger than 1,
it not necessarily obvious how to achieve this and that question merits more attention.

In the following we denote $z\equiv (z_0\cdots, z_N)^T$ in $\mathbb{C}^{N+1}$. Similarly
$a$ is in $\mathbb{C}^{N+1}$.

\begin{lem}
Suppose in the system of Proposition \ref{prop:Galilean Invariance} agent ``0"
is the only leader and its orbit is given by $z_0(t)=e^{i\omega t}$.
Then as $t\rightarrow \infty$ the orbit of the system converges to
\bsenn
z(t)= a e^{i\omega t} \quad ,
\esenn
where Galilean Invariance implies that $a=(a_0\cdots,a_N)\in \mathbb{C}^n$ satisfies:
\bsenn
-\omega^2 a = (f L_\rho +i\omega g L_r)\,a \quad .
\esenn
And in particular $\lim_{\omega\rightarrow 0}\, a_k(\omega)=1$ for all $k\in\{0\cdots,N\}$.
\label{lem: perturbations dont die}
\end{lem}

This Lemma is immediately relevant to the present discussion because it
indicates that certain low-frequency perturbations in the orbit of the leader
\emph{cannot possibly die out}: it is impossible to design a decentralized
system where all perturbations die out.
The situation is in fact probably worse: as the size of the system grows,
it seems that, say, $\max_{\omega\in\mathbb{R}^+}|a_N(\omega)|$ must also grow,
probably with some power $1/d$ of the number $N$ of agents, where perhaps $d$
is related to a notion of the dimensionality of the ``communication graph".
This question has received some attention (\cite{PSH}), but not nearly enough.

The trouble doesn't even end there. When we study flocks --- whether biological
or technological --- it is natural to assume that each
agent performs the same computation to determine its acceleration. When you imagine
a large flock then for all agents `sort of' in the middle of that flock that
assumption is reasonable: all agents are hardwired equal. We'll call this
\emph{homogeneity}. It is hard to give a formal definition of this concept since
it cannot be completely true for a finite flock moving in $\mathbb{R}^d$:
the agents at the physical boundary of the flock necessarily have less neighbors.
Since perturbations cannot all die out --- as we just showed,
the boundary conditions may well influence the dynamics of the flock.
We can see two possibilities: either boundary conditions (or some natural
subset thereof) don't matter, or else there is a smart and natural way of
uniquely specifying them. To the best of our knowledge this important question has
hardly if ever been addressed (except \cite{WS}, unpublished)!

\begin{centering}\section{Decentralized Flocks in a Medium, Mass} \label{chap:four}
\end{centering}
\setcounter{figure}{0} \setcounter{equation}{0}

We do not live in a vacuum!
Whatever our method of locomotion is, in a medium we
can perceive our \emph{speed} with respect to the medium, by the effort we expend
to maintain the speed to overcome friction. (Though without a compass we cannot
measure the direction of that speed.)
The consequence of perceiving speed is that there is a preferred direction
for the agents, namely: forward (in the direction of the velocity).
Thus flocks in dimension 2 can orient themselves so that their desired
configuration is aligned along the velocity of the flock.

To formulate those equations in $\mathbb{R}^2$ let's go back to the system
given in Proposition
\ref{prop:Galilean Invariance} in the coordinates of the
agents given by $x_k$ where $x_k=z_k+h_k$ (the first of the two equations).
We now interpret $x_k$ and $h_k$ as vectors in $\mathbb{R}^2$.
These equations are not truly decentralized anymore! An agent unsure where North is
can only measure $|h_k-h_{k-1}|$ (ie: relative distances). However flying in a medium,
the agent can perceive their direction of flight and orient itself
according to it (as long as their speed is non-zero.)
Thus the rotational symmetry of the truly decentralized equations
in dimension 2 can be broken. We argue that this is how geese manage
to fly in formation. Here are the equations:

\vskip .1in
\begin{prop} (\cite{flocks2}) Let $\theta_k$ be angle that $\dot x_k$ makes
with the positive $x$-axis. Denote by $R_\theta$ the rotation by $\theta$.
The equation of motion for the \emph{orientable} flock is:
\bsenn
 \ddot x_k = f \sum_{i\in{\cal N}_\rho(k)}\,L_{\rho,ki} (x_{i}- R_{\theta_k} h_i) +
                    g \sum_{i\in{\cal N}_r(k)}\,L_{r,ki} \dot x_{i} \quad ,
 \esenn
 as long as all $\dot x_i$ are non-zero. (The equation has a singularity at
 $\dot x_i=0$.)
\label{prop:orientable}
\end{prop}

\begin{figure}[pbth]
\center
\includegraphics[width=3.5in]{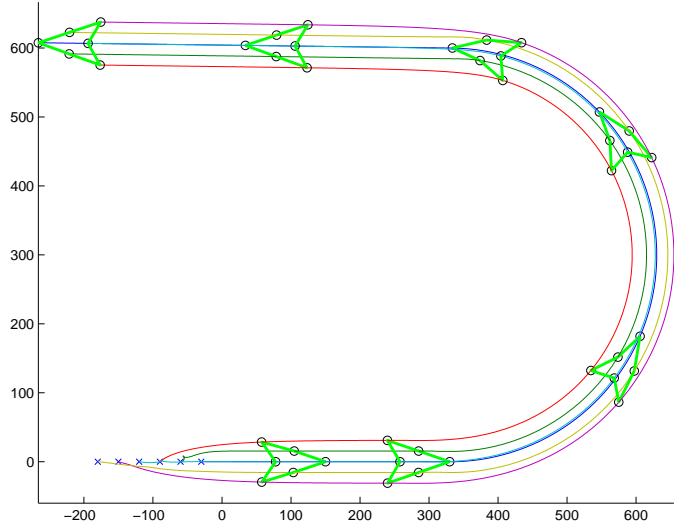}
\caption{(color online) \emph{This figure is a simulation of a flock
turning around using
Definition \ref{prop:orientable}. Note that the
orientation of the flock's configuration changes. Each `christmas
tree' shaped hexagon is a snapshot of the position of the flock,
the lines that form the figure only facilitate visual inspection,
they have no physical or mathematical content.}}
\label{flocks2fig1}
\end{figure}

In Figure \ref{flocks2fig1} taken from \cite{flocks2} we show a numerical simulation
using the Equations of Proposition \ref{prop:orientable}. The figure shows
that in principle it is possible to re-orient a
(mathematical) flock as it turns.
(Remark: The way we did it in that paper is not exactly how we propose
to study that question here. But it does provide a proof of concept.)
(See also \cite{flocks3})
The important question is: how can we understand the re-alignment
of a flock in 2 (or 3) dimensions that changes its
direction when certain leaders at its boundary initiate
a course change. We invite the reader to study these equations!
The non-linearity makes these equations a real challenge to study
(a few results are given in \cite{flocks2}).

This leaves two other questions. It is possible to introduce a desired
speed $V$ in the above equations. Furthermore in a realistic flock it is
likely that the masses $m_k$ of the agents have a distribution with average
1 (as opposed to being exactly 1). The equations (in the
absence of any external forces) then become:
\bse
 m_k \ddot x_k = f \sum_{i\in{\cal N}_\rho(k)}\,L_{\rho,ki} (x_{i}- R_{\theta_k} h_i) +
                    g \sum_{i\in{\cal N}_r(k)}\,L_{r,ki} \dot x_{i}
                    + \alpha \left(1-\frac{V}{|\dot x_k|}\right)\dot x_k \quad .
\label{eq:full-equation}
\ese
Here $\alpha <0$. We expect that both the presence of desired
speed (a cruise speed) and the variation in the masses act to stabilize
the equations to some extent. In a realistic setting we probably \emph{want}
flocks to be able to travel at different speeds: in different
environments (say, cities or freeways) the desired cruise speeds should
be different. Furthermore even when we agree on a single
desired cruise speed, it appears unrealistic that
individual agents in dense traffic can be be programmed to determine
their absolute speed accurately enough so as to avoid collisions with their immediate neighbors.
Nonetheless we offer Equation \ref{eq:full-equation} as a possible starting point
for a more general mathematical study of flocking in 2 dimensions than
the equation in Proposition \ref{prop:orientable}.

\begin{centering}\section{Symmetry and Stability} \label{chap:five}\end{centering}
\setcounter{figure}{0} \setcounter{equation}{0}

Now we come to what is certainly one of the most intriguing
and important questions concerning flocks:
What happens if a member at the ``boundary" of the flock suddenly its velocity
for whatever reason (a predator is observed or a traffic light changes etc)?
How is the capability of following the new --- supposedly more advantageous ---
affected by the size of the flock? The (in)ability to effectively follow
the new course \emph{should} limit the maximum allowable size of the flock.
What is that size, and how can we implement that in a `technological' flock?
Recall that we assume that a flock is asymptotically stable, so that no matter
what the course change is, the other agents will ultimately follow the leader.
The catch is that if, prior to stabilizing, fluctuations become too large: the flock is still
not viable. Sometimes this is informally referred to as `transient stability'.
In the literature the concept of `string stability' has played an important role
(\cite{SH1}, \cite{SH2}).

Let's look at the Standard Example (Definition \ref{def:standard example}).
That system is given in such a way that $\lim_{t\rightarrow \infty}\,z(t)=0$
and $\lim_{t\rightarrow \infty}\,\dot z(t)=0$. This is important in what follows.
Consider the quantities $\max_k \max_{\omega\geq 0}\,|a_k(\omega)|$
and of $\max_k \max_{t\geq 0} |z_k(t)|$.
It turns out that the maxima are achieved for $k=N$, which simplifies the notation.

\vskip .1in
\begin{defn} We call the system `harmonically unstable' if
\bsenn
  \dfrac1N\, \ln \max_{\omega\geq 0}\,|a_N(\omega)| >0 \quad ,
\esenn
and `impulse unstable' if
\bsenn
  \dfrac1N\, \ln \max_{t\geq 0}\,|z_N(t)| >0 \quad .
\esenn
The system is `flock unstable' if either holds, `flock stable' otherwise.
\label{def:stable}
\end{defn}

\begin{theo} (\cite{flocks5}, \cite{flocks6}, \cite{flocks7}, \cite{flocks67}) The system given in Definition
\ref{def:standard example} with $r=\rho$ is harmonically and impulse stable
$\rho=1/2$ and harmonically and impulse unstable for $\rho\in (0,1)\backslash \{1/2\}$.
\label{thm: stable}
\end{theo}

There are several problems with this. To set up the definition of flock stability
more generally, one would have to rely on `homogeneity' which we don't know how to define
(see Section \ref{chap:one}). The second problem is that the Theorem is so hard to prove that it seems unlikely that substantial generalizations can be made
--- at least not using the methods of those papers.

\begin{figure}[pbth]
\center
\includegraphics[width=3.5in]{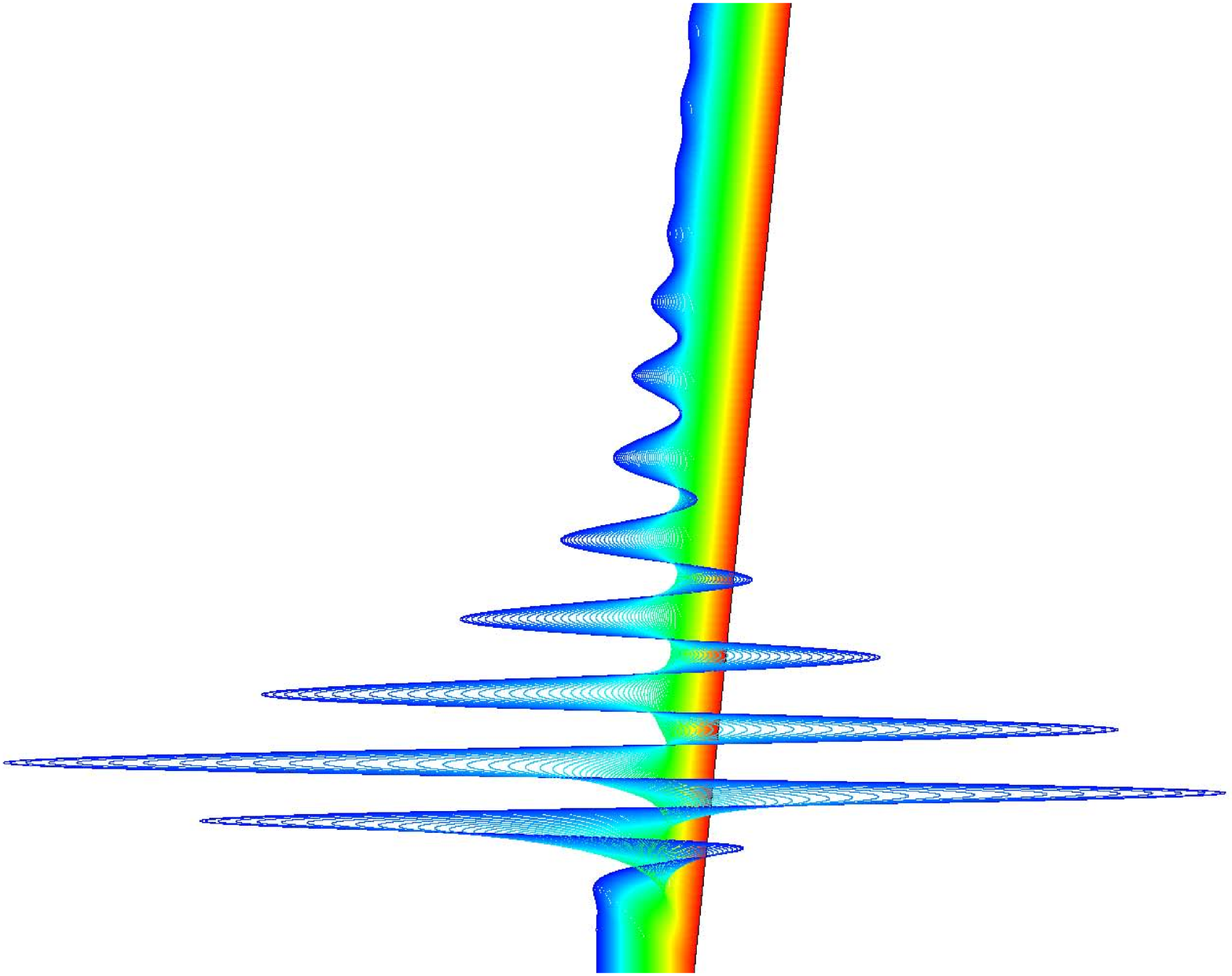}
\includegraphics[width=1.5in]{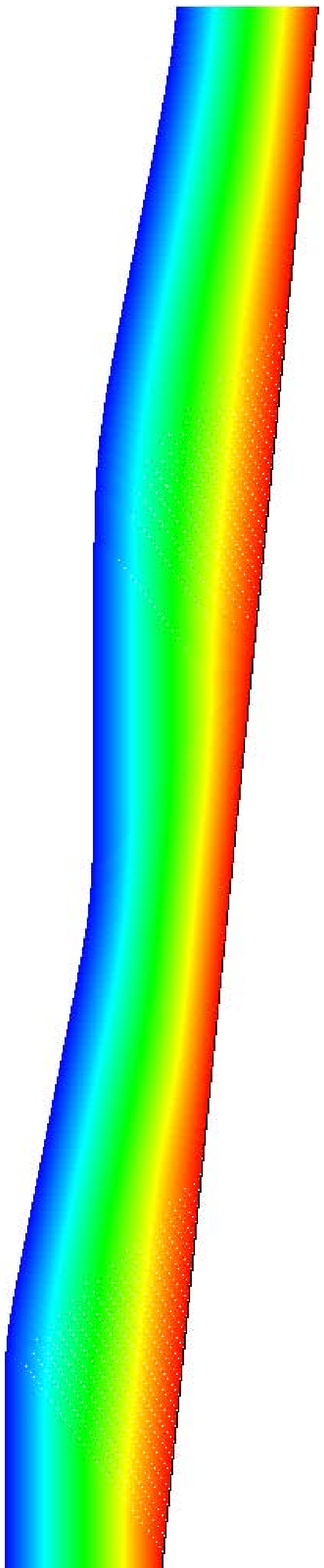}
\includegraphics[width=1.44in]{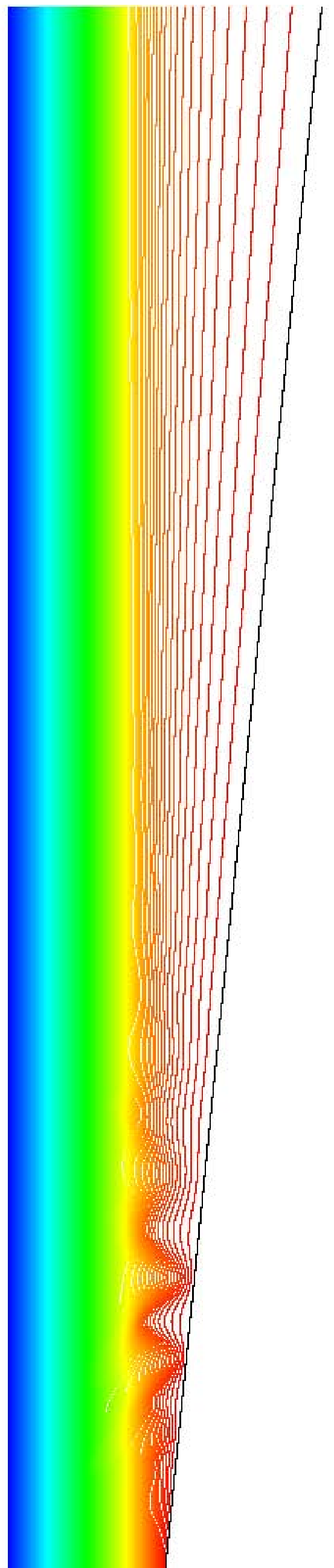}
\caption{(color online) \emph{Numerical solution of the equation of motion of
Definition \ref{def:standard example} with $r=\rho$, $f=-1$, $g=-2$,
$N=100$. Time $t$ is vertical
and position ($x$) is horizontal, each Figure being approximately a 1000 by 1000
square. Each agent initially has unit distance to its
neighbors, so that the flock initially is 100 units wide (horizontally).
At $t=0$ the rightmost agent, the leader, starts moving with constant velocity 0.1 to the right.
Simulations for $\rho=0.45$ (left), $\rho=0.50$
(middle), and $\rho=0.55$ (right). }}
\label{fig:TIME}
\end{figure}

\begin{figure}[pbth]
\center
\includegraphics[width=2.2in]{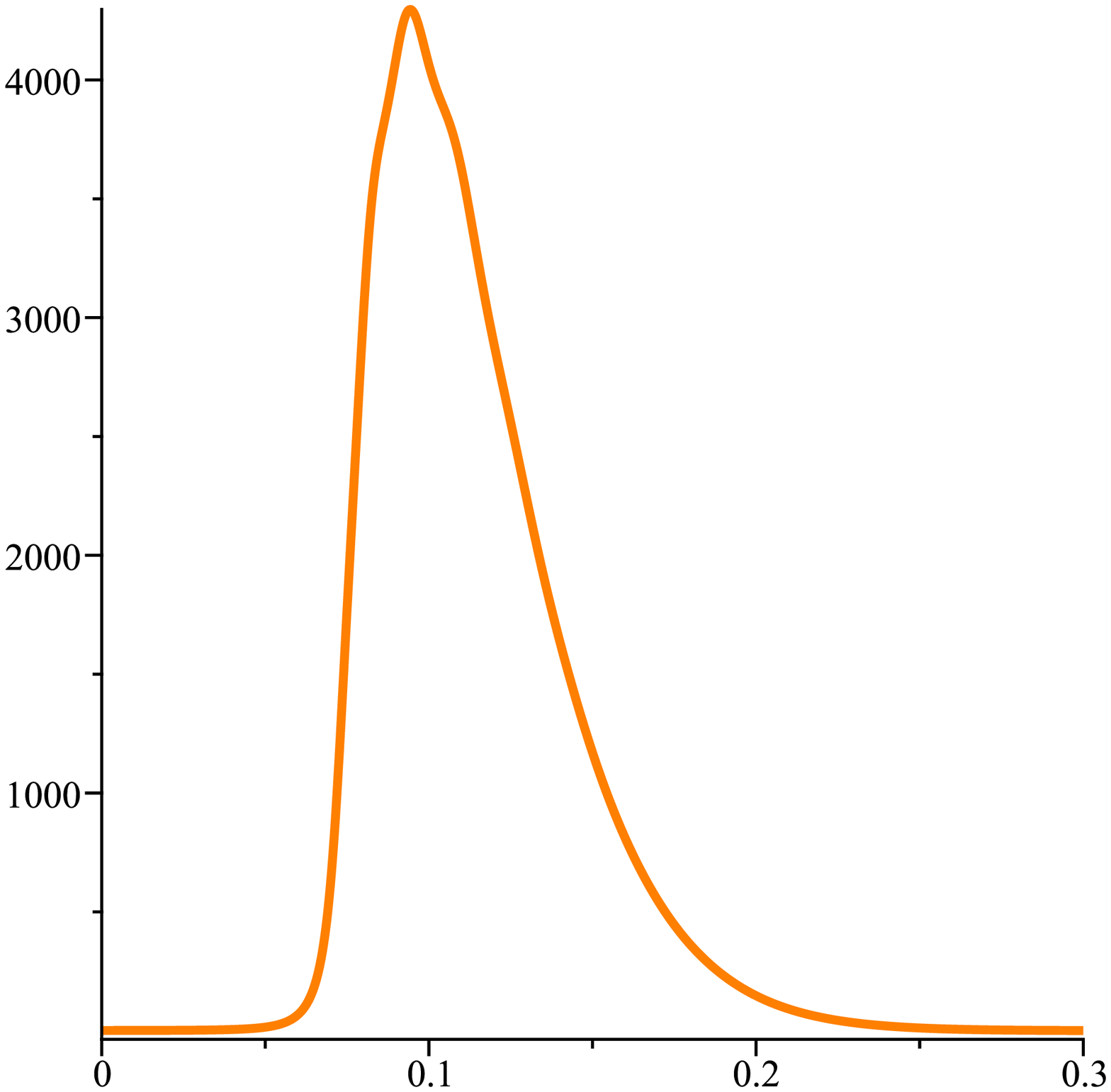}
\includegraphics[width=2.2in]{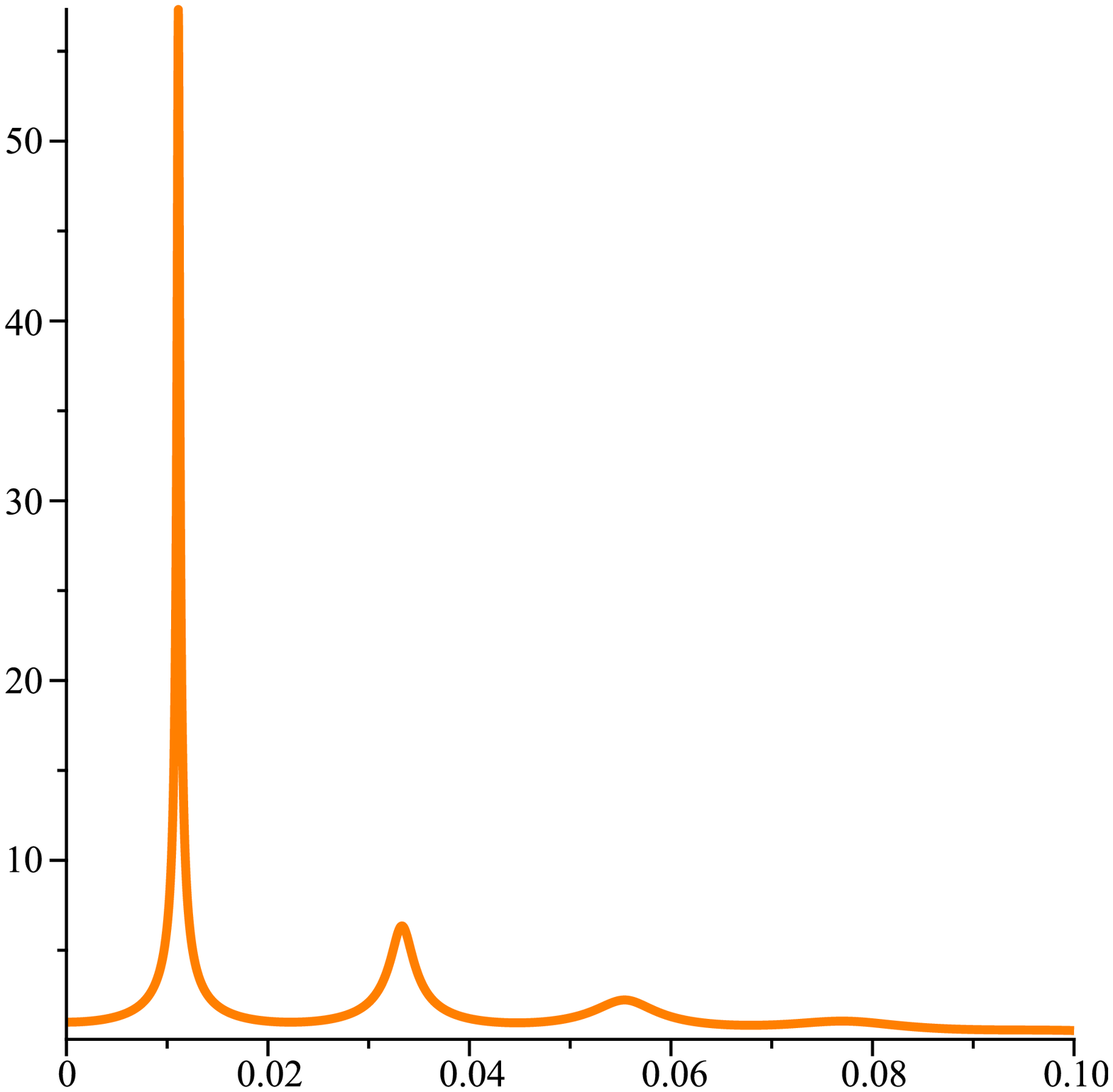}
\includegraphics[width=2.2in]{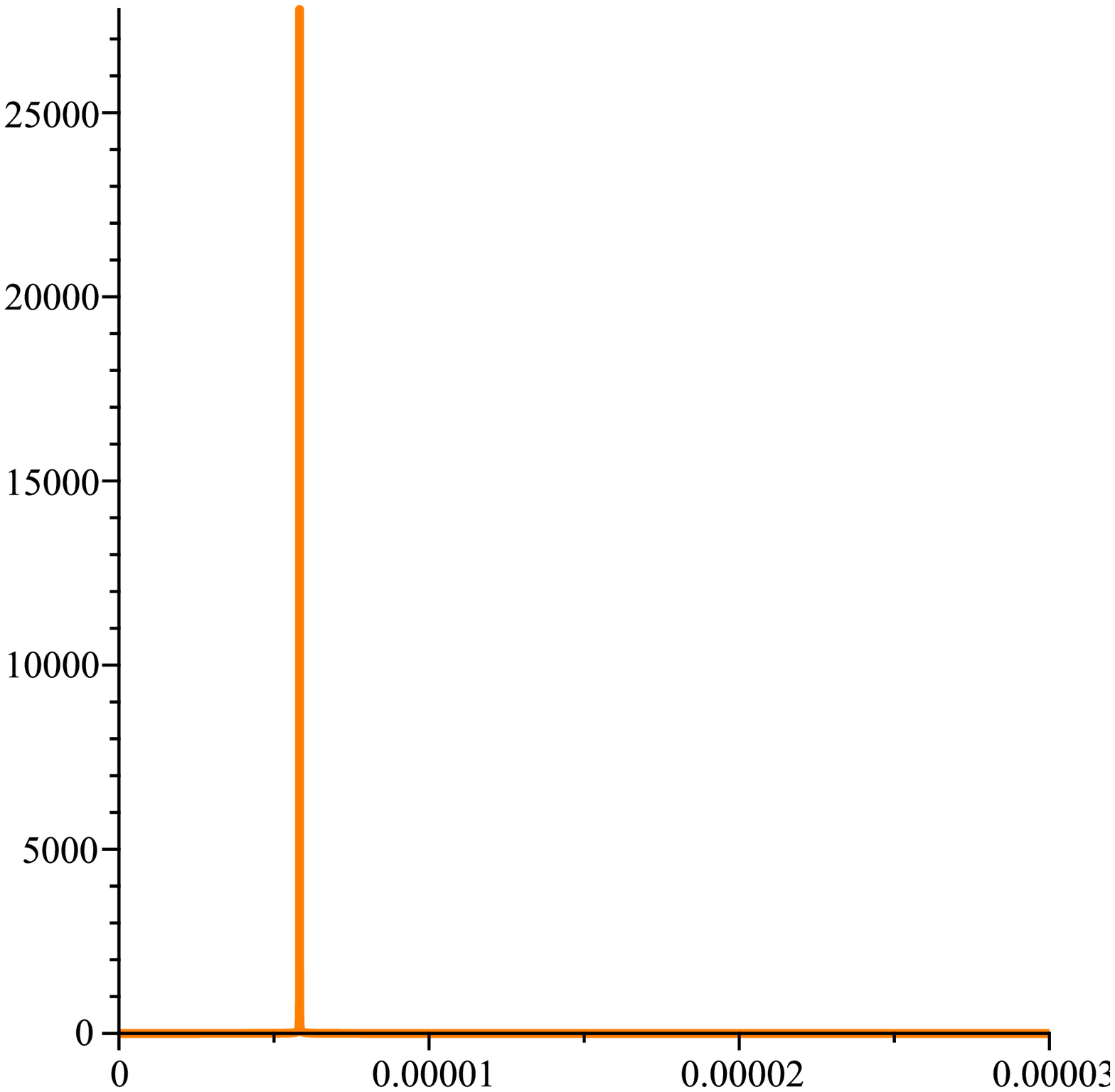}
\caption{(color online) \emph{The frequency response function for the system
of Figure \ref{fig:TIME}. The absolute value of $a_N(\omega)$ for $\rho=0.45$ (left),
$\rho=0.50$ (middle), and $\rho=0.55$ (right). Observe the different
scales in these Figures!}}
\label{fig:absolute}
\end{figure}

\begin{figure}[pbth]
\center
\includegraphics[width=2.2in,height=0.8in]{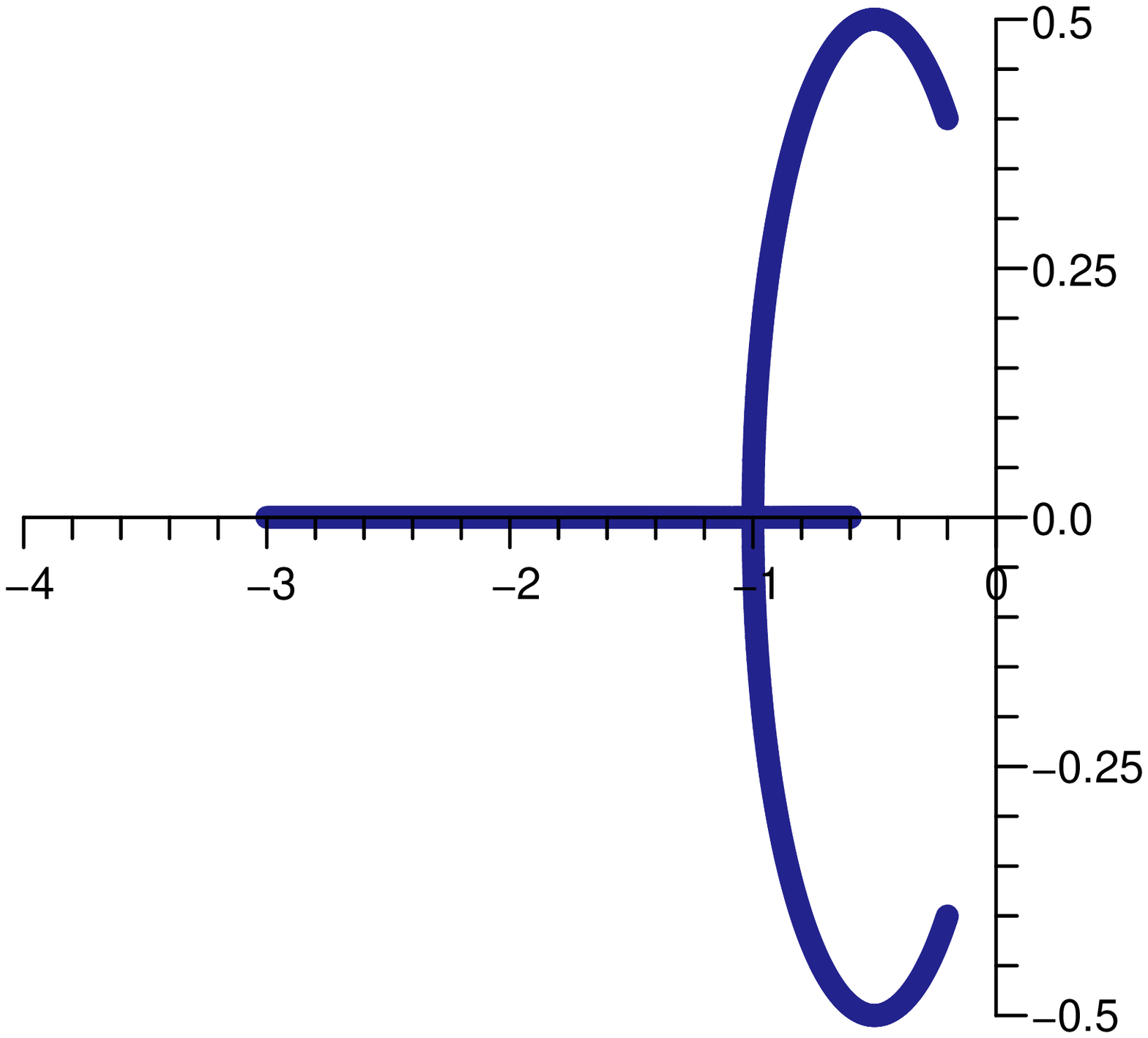} $\quad$
\includegraphics[width=2.2in,height=0.8in]{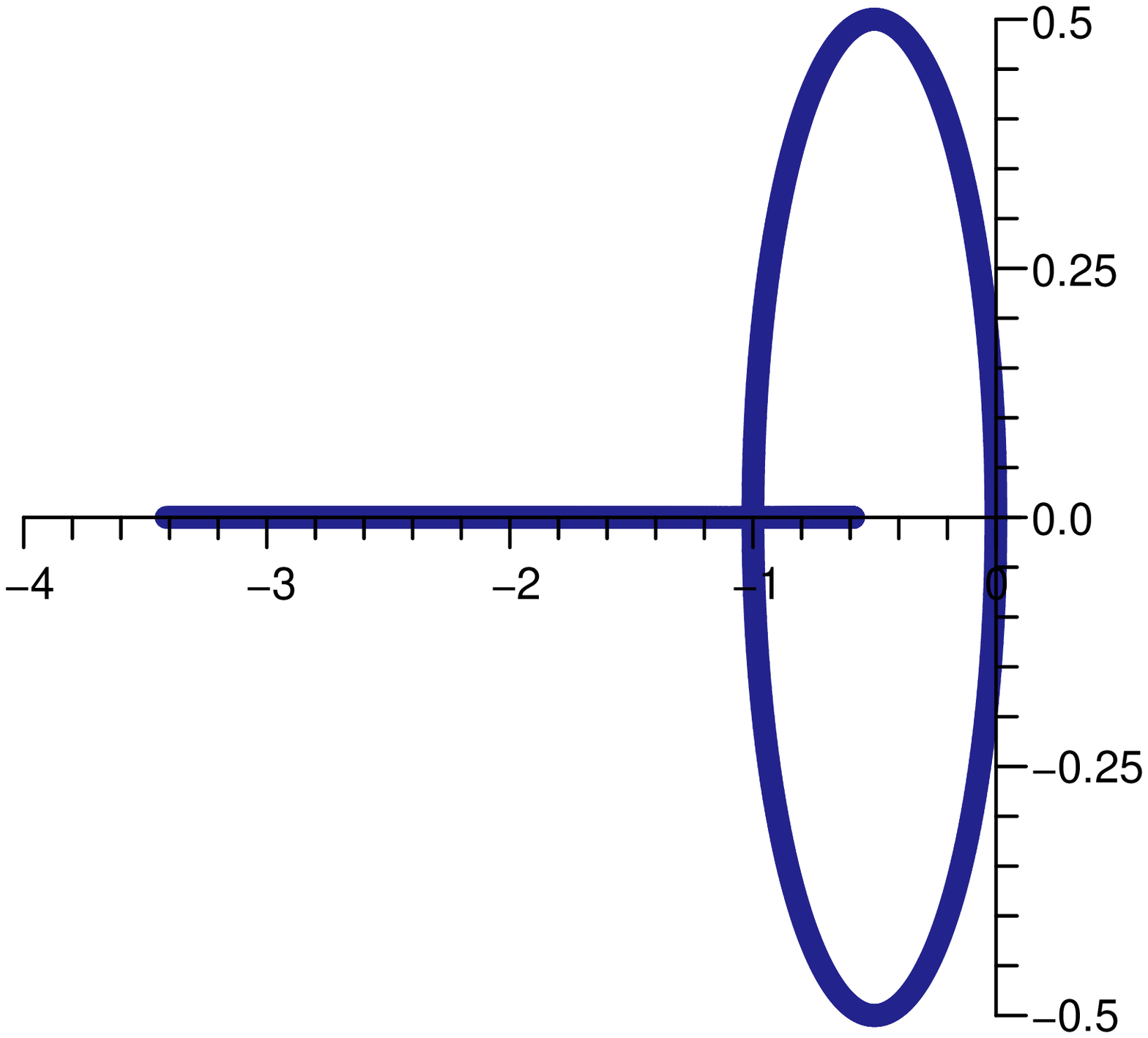} $\quad$
\includegraphics[width=2.2in,height=0.8in]{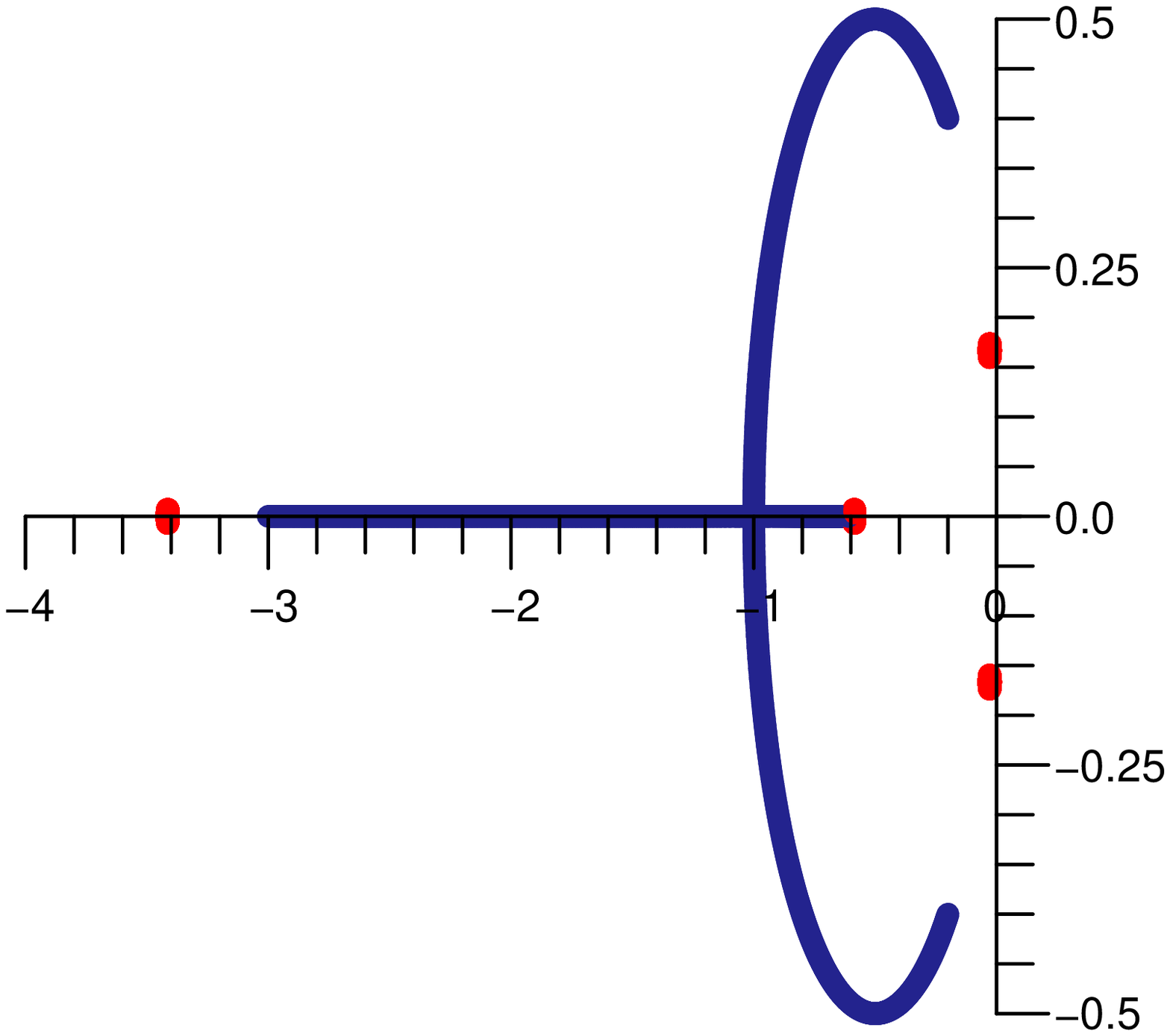}
\caption{(color online) \emph{Schematic representation in the complex plane
of the eigenvalues associated to the first order ODE of the system given in
Figure \ref{fig:TIME}. For $\rho<0.5$ the real part of the eigenvalues are less than a
negative constant (right). When $\rho$ approaches 0.5, the eigenvalues approximate
the imaginary axis and the origin (middle). When $\rho>0.5$ the process reverses, except that two pairs of eigenvalues (in red)
 is spun off. One of these is exponentially (in $N$) close to zero (left).}}
\label{fig:eigenvalues}
\end{figure}

The effect of changing $\rho$ are truly dramatic as the accompanying figures show.
(This was also observed in \cite{BH}.)
In Figure \ref{fig:TIME} (taken from \cite{flocks67}) we show a simulation of the system of Definition
\ref{def:standard example} with  $r=\rho$ and with 100 agents, and only slightly vary $\rho$: 0.45, 0.50, and
0.55. For $\rho<1/2$ wild fluctuations of relatively high frequency occur wiping out
every semblance of cohesion in short order. For $\rho>1/2$ a very large fluctuation
occurs over a very large time-scale (a factor $10^3$ longer than we can exhibit here).
As indicated before, both systems ultimately stabilize: in practice this is moot,
because the transient fluctuations would have destroyed cohesion of any
actual physical flock already.
In Figure \ref{fig:absolute} we show (the absolute value of) the frequency response
functions in the three cases: $\rho=0.45$, $\rho=1/2$, and $rho=0.55$. Note the
difference in scales!
When $\rho=1/2$ the spectrum the frequency response functions
have a very peculiar form that enables us to deduce the actual
dynamics of the individuals. They undergo a damped stop-and-go motion (analyzed
in \cite{flocks5} and reminiscent of post-holiday traffic into urban areas).

These results reveal yet another curious fact. In finite-dimensional (linear) systems
we are used to the intuition that large fluctuations occur when the
eigenvalues associated with the first order ODE
(the poles of $a_N(\omega)$) get close to the imaginary axis.
Here we vary the dimension $N$ of the system and now \emph{this intuition turns out
to be completely false}! In this case we can calculate the eigenvalues of the system
(see \cite{tridiagonal} and \cite{flocks67}). The results are displayed in Figure
\ref{fig:eigenvalues}. When $\rho>1/2$ there is an eigenvalue (indicated in red)
extremely close (exponentially in $N$) to 0; that eigenvalue dominates the behavior
of the system. When $\rho=1/2$ there are a couple of eigenvalues roughly at distance
$N^{-2}$ of the imaginary axis; these few also dominate the dynamics. However
when $\rho<1/2$ \emph{all eigenvalues are bounded away from the imaginary axis,
uniformly in $N$}! Now all eigenvalues contribute to the dynamics and this
gives surprising results: very large fluctuations. From the point of view of
studying the eigenvalues, this is completely counter-intuitive.

In view of these difficulties we close by encouraging (again) the search for
a more general point of view.
Let's go back to the more general system given in Proposition
\ref{prop:Galilean Invariance}.
The physical ``work" $W_k$ done by the forces acting upon agent $k$ between
time 0 and time $t$ is given by $\int_0^t\, \ddot z_k \, \dot z_k\, dt$ which
equals $\frac12 (\dot z_k^2(t)- \dot z_k^2(0))$. Employing $(\cdot,\cdot)$,
the usual Hermitian inner product on $\mathbb{C}^{N+1}$, the sum of the $W_k$'s can
be written as $\frac12 ((\dot z,\dot z)(t)- (\dot z,\dot z)(0))$. After some algebra
one obtains \emph{for arbitrary Laplacians} $L_rho$ and $L_r$:

\begin{prop} The equations given in Proposition \ref{prop:Galilean Invariance}
imply:
\bsenn
\dfrac12 \big[ (\dot z, \dot z)(t)-f\, (L_\rho^S z,z)(t)\big]=
\dfrac 12 \big[(\dot z, \dot z)(0)-f\, (L_\rho^S z,z)(0)\big]
+g \, \int_0^t\, (L_r^S\dot z, \dot z)\, dt + f\,\int_0^t\,(L_\rho^A z,\dot z)\, dt \quad .
\esenn
Here for a matrix $L$ we have written $L=L^S+L^A$, the latter two denoting the
symmetric and the anti-symmetric parts.
\label{prop:integration}
\end{prop}

Specialize to the Standard Example. One easily sees that
$(\dot z, \dot z)(0)=N$ while $(L_\rho^S z,z)(0)=0$.
Perhaps somewhat surprisingly, the matrices $L\rho^S$ and $L_r^S$ have
one negative eigenvalue (due to the special form of the equation of
motion of the leader). This eigenvalue appears to have small modulus.
Similarly, the matrix $L_\rho^A$ has non-zero entries only near the
`boundary-terms'. Thus the left-hand side tries to be semi-definite positive
while the last two terms of right-hand side is ``try" to be negative, pushing
the RHS to 0 to zero, but boundary effects slow that down.
A more careful analysis would be extremely interesting. The new aspect
is that Proposition \ref{prop:integration} is valid is completely independent
of the topology of the network.

\begin{centering}\section{Conclusion} \label{chap:six}\end{centering}
\setcounter{figure}{0} \setcounter{equation}{0}

The first point we argue in this position paper is that considerations
of a fundamental nature (symmetry, friction) can help lead us to formulate
the simplest possible equations governing the motion of de-centralized flocks.
In one dimension these are linear (Proposition \ref{prop:Galilean Invariance}),
but in higher dimension that is not the case (Section \ref{chap:four}).
We contend that these equations (especially the ones in dimension 2 or 3) have been
insufficiently studied in favor of ad hoc models.
For example it is not well-known enough that perturbations cannot all die out as they travel through the flock (Lemma \ref{lem: perturbations dont die}. Neither do we
know of any detailed studies of flocks in $\mathbb{R}^2$ or $\mathbb{R}^3$ that change direction and orientation s suggested in Section \ref{chap:four}.

The second point we bring up is that of flock-stability. In Definition
\ref{def:stable} we propose that the stability of a flock is determined by
how certain response functions grow with the size of the (homogeneous) flock.
A fundamental result, Theorem \ref{thm: stable}, states flock stability
is intimately related to internal symmetry of the flock.
This result together with Figure \ref{fig:eigenvalues} shows that it is
completely misleading to
look at the location of the eigenvalues for these linear systems: one must
look at the time domain as well (the impulse response function
of Definition \ref{def:stable}). This is not widely acknowledged in the literature.
Since time-domain results as those of Theorem \ref{thm: stable}) are
so hard to obtain, the need for new methods --- preferably based on general
principles --- is great. In Proposition \ref{prop:integration}
we propose to gain some insight in a general setting by evaluating the physical
``work" done by the system.

\section*{\uppercase{Acknowledgements}}
\noindent
I am grateful to Borko Stosic for numerous discussions.

\vspace{\fill}

\end{document}